# Terahertz Microscopy Through Complex Media

Vivek Kumar[1,3], Vittorio Cecconi[1,2], Antonio Cutrona[1,2], Luke Peters[1,2], Luana Olivieri[1,2], Juan S. Totero Gongora[1,2], Alessia Pasquazi[1,2], Marco Peccianti[1,2]*

[1]Emergent Photonics Lab (EPic), Department of Physics and Astronomy, University of Sussex, BN1 9QH, UK.
[2]Emergent Photonics Research Centre, Department of Physics, School of Science, Loughborough University, LE11 3TU, UK.
[3]Laboratoire Kastler Brossel, ENS-Universite PSL, CNRS, Sorbonne Universite, College de France, 24 rue Lhomond, 75005 Paris, France.
*m.peccianti@lboro.ac.uk

**Abstract:**
Manipulating broadband fields in scattering media is a modern challenge across photonics and other wave domains. Recent studies have shown that complex propagation in scattering media can be harnessed to manipulate broadband light wave packets in space-time for focusing, imaging, and computing applications. Interestingly, while many proposed methodologies operate on intensity-based assessment of scattered fields, often in the spectral domain, from a pure transmission-function perspective, scattering operates as a linear field-level combinatory process, i.e., the superposition of transformation of unit excitations. As a result, we recently demonstrated that gaining experimental access to instantaneous scattered fields, as available through time-domain spectroscopy in the terahertz spectral range, in conjunction with sparse light excitation typical of ghost imaging, provides a key advantage in enabling the functionalisation of scattering, exposing a novel modelling paradigm. In this paper, we provide experimental proof of reconstructing 1-dimensional object features through a scattering medium using a fully broadband time-domain terahertz approach.

**Introduction**

Inhomogeneous complex media, such as biological tissues or atmospheric turbulence, exhibit wave scattering, typically viewed as an inevitable perturbation or a nuisance. Due to the recurrence of scattering and interference, this phenomenon seemingly obliterates both the spatial and time information of any transmitted wave[1,2]. As commonly experienced, scattering ultimately limits imaging, a challenge so fundamental that it is the sole focus of a large multi-domain research effort. However, multiple scattering is a highly complex but deterministic process; hence, information is scrambled but not completely lost. Wave propagation modelling in scattering media is an invaluable field of research and is regarded as a key objective in free-space communication and biomedical imaging[3–5]. Significant advancements have addressed wave control through disordered materials, resulting in scattering-assisted images using wavefront shaping techniques[6–9]. Within this research area, wavefront shaping methods have been explored in high-resolution imaging modalities such as optical coherence tomography[10], fluorescence microscopy[11], two-photon microscopy[12,13], photoacoustic microscopy[14] and improved resolution of images. These imaging modalities demonstrated that controlling the propagation of scattered light can be essentially addressed by optimising an input wavefront, a process that can be deployed via feedback optimisation[15,16], phase conjugation[17,18] or by measuring the optical transfer matrix[19–22].

In its most basic embodiment, the deterministic approach towards space-time wave synthesis in scattering media is pivoted onto the idea of probing the media with spatially orthogonal illuminations and sensing their corresponding scattered wave at the output, with the aim t. For a sufficiently complex medium, this means that the spatiotemporal field pattern at the output can be associated with each input sampling function (or pattern) without ambiguity, as the transformation between input and output can be approximated as unique -although time-reversal would require the observation of the scattered wave in all directions, including the backscattering at the input. Via the application of the theoretical foundation presented in ref[23,24], scattering can be managed as a linear combinatory process. Once a spatiotemporal output set is recorded for a series of orthogonal input illumination, we can project an arbitrary desired scattered field onto this set, determining the required input excitation. Indeed, this is not a statistically founded concept, and it is not based on any specific portion of the scattered field (i.e., ballistic or diffusing).

Trivially, in this vision, direct access to the instantaneous scattered fields is essential from an experimental point of view, as the superposition principle does not generally apply to photonic intensity-based scattering assessments. Additionally, in state of the art, the extension of this concept to broadband illumination is challenged by the fact that the combinatory scattering element is, in general, a function of the frequency, while this aspect is marginally relevant when the superposition of scattered spatiotemporal waveforms is considered. Modern THz-TDS (terahertz time-domain spectroscopy) enables a direct route to measure the electric field oscillations in broadband pulses. Therefore, an intriguing question is whether working with the spatiotemporal electric field and the prospect of field-sensitive detection may shed light on reconstructing the combinatory scattering element of a sample on a broadband spectrum. In such a direction, we recently demonstrated field-level THz wave synthesis[25].

The core research literature in scattering functionalisation in photonics has presented a significant use of off-the-shelves spatial light modulators. Interestingly, spatial light modulation is somewhat challenging in the Terahertz domain[26]. The spatial modulation of THz waves has predominantly relied on indirect methods, including the use of spinning disk masks[27,28], optically controlled carrier-based masking in silicon[29–32], meta-material spatial light modulators[33–35], spatial encoding of optical probe beams in electro-optic imaging systems[36] as well as optical pump beams in systems using nonlinear[37–40] or spintronic THz emitters[41]. Many of those wavefront modulation schemes are at the basis of single-pixel ghost imaging and other forms of imaging analysis. The use of THz spatial light modulation for achieving comprehensive wavefront control through scattering media remains at an embryonic stage. We can also observe that in this context, while typical illuminated areas in THz experimental embodiments recall similar implementations in optics, the spatial density of independent input modes in a scattering media is several orders of magnitude lower. This means that available diffraction-limited spatial light modulation cannot access a significant diversity of propagating modes in scattering samples[42]. As a possible solution implemented here, a near-field coupled knife edge can be used to modulate the spatial properties of an input light field exposing modes that are not coupled with input radiative fields. Here, we demonstrate the 1D imaging through scattering via a THz-TDS imager to decompose the time-domain scattered field at the output facet. The spatiotemporally resolved field-based measurements correspond to the set of spatially modulated THz wavefronts. The modulation set is simply obtained by scanning the input facet of the scattered 1D knife-edge placed in the near field.

**Results and Discussion**
**Spatiotemporal THz field mixing in Scattering Media**
We begin addressing how fields at source and detection planes are connected by a complex combinatory matrix element of a scatterer precisely for the case of broadband pulse illumination. We define the input-output field relationship via the space-time impulse response of scattering media, $T_x(x, y; x', y'; t, t')$, as

$$E_d(x', y', z_d; t') = \iiint T_x(x, y; x', y'; t, t') E_s(x, y, z_s; t) dx dy dt \qquad (1)$$

where, $E_d(E_s)$ is spatiotemporal field distribution at the detection/source plane in the cartesian coordinates defined by $(x', y', z_d)/(x, y, z_s)$, respectively. In frequency domain $\omega$, equation (1) reads as follows,

$$\tilde{E}_d(x', y', z_d; \omega) = \iint \tilde{T}_x(x, y; x', y'; \omega) \tilde{E}_s(x, y, z_s; \omega) dx dy \qquad (2)$$

where $\tilde{E}_d$, $\tilde{E}_s$ denote the time-Fourier transform of the fields at detection, source plane and $\tilde{T}_x(x, y; x', y'; \omega)$ represents coherent transfer function for a scattering sample.

THz source providing an ultrafast broadband pulse with a duration shorter than Thouless time (the typical time scale over which light diffuses across the medium ) results in randomised field transient in both space and time after propagating through scattering media[43,44]. In figure 1, we recorded the transmitted time-resolved THz field distribution in response to Gaussian beam illumination on the scattering sample (sample details are in supplementary S1), as shown in the schematic (figure 1a). As expected, scattering introduces a time-dependent morphology of the scattered field profile. Specifically, the THz field distribution resulted in complex spatial interference patterns in the transmission (figure 1b). The evolution of the spatial field on an ultrafast time scale shows the transition from the ballistic to the diffusive scattering regimes. Initial spatial THz field profiles at 8.85 ps and 9.45 ps indicate a ballistic propagation of THz waves associated with a low statistical occurrence of photon scattering, i.e., the THz field maintains its spatial momentum content. Later in time, the diffused photons appear, which is entirely randomised as a result of several scattering occurrences inside the medium, as shown by the spatial frames at 10.35 ps, 14.4 ps, 15 ps and 18.9 ps. Also, the scattering process becomes more apparent in the temporal progression of distinct pixels that constitute the considerably broadened pulses in conjunction with the different time delays. However, the overall observed time delay in the pulses can be attributed to the thickness of the scattering sample (5.53 mm). As a result of the dispersive characteristics inherently present in scattering media, the temporal widening of a scattered pulse occurs concomitantly with the attenuation of its peak values. The primary cause of attenuation in the scattered pulses is attributed to the scattering of high-frequency components of the impinging pulse, as illustrated in figure 1c. The measured values of the mean free path ($l_s$) for the sample are shown in figure 1d. In our experiments, we observed the variation of $l_s(\omega)$ ranging between 1 mm to 10 mm within the source bandwidth. Note that because the scattering media is near field coupled with the detector, scattering can appear highly inhomogeneous in spectral properties at the output. The spatial spectral profiles of the THz field at 0.3-0.7 (figure 1e). THz exhibits the formation of broadband speckles, characterised by subwavelength spot size and a disordered phase distribution. We highlight that the detection is near field coupled with the scattering media, hence the resolution is not bounded by the wavelength.

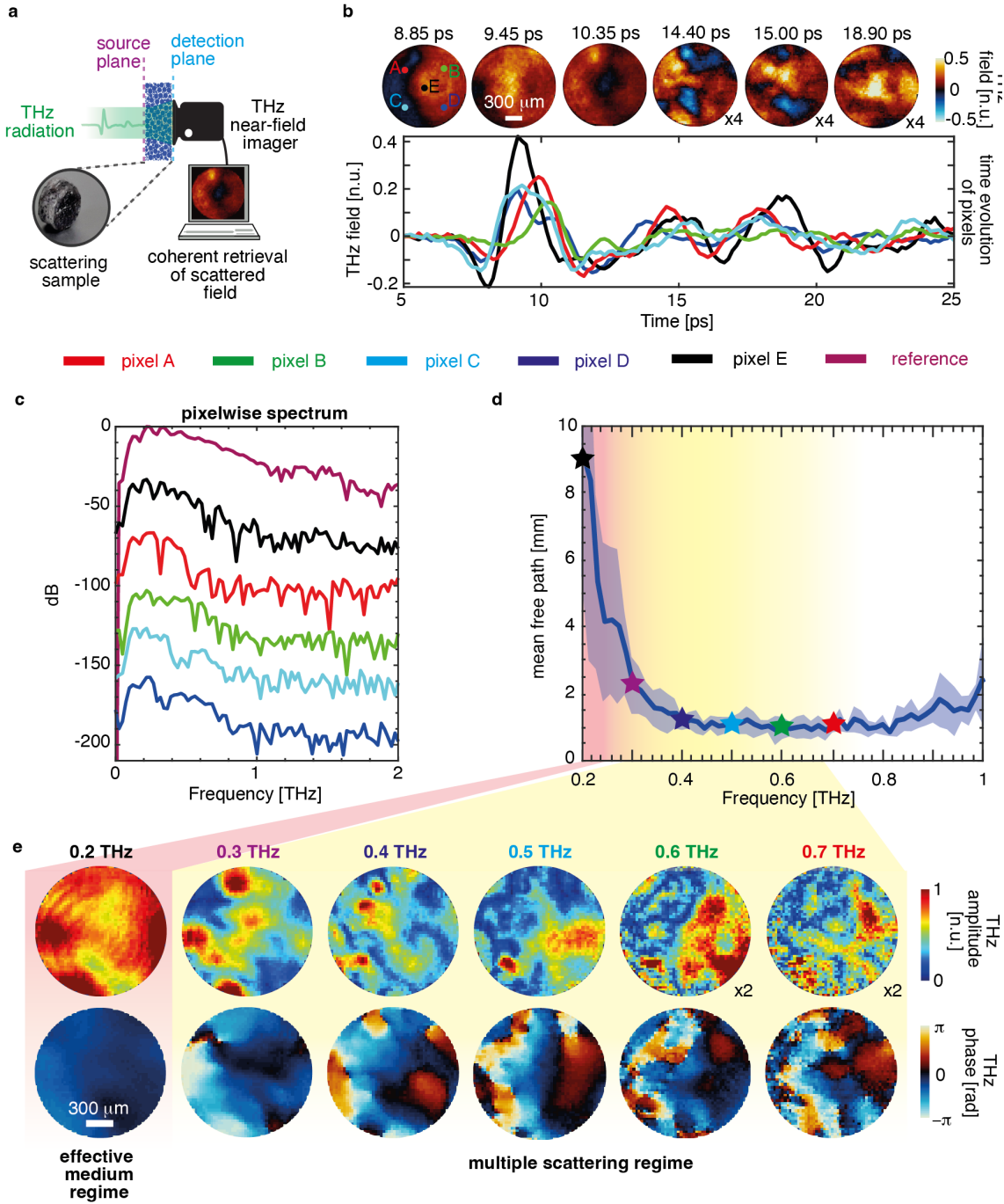

**Figure 1: Broadband speckle formation of THz fields. a.** Schematic showing the projection of a pulsed THz field at the source plane of the scattering sample and the collection of the scattered field via a time-domain terahertz imager. **b.** Spatiotemporal THz field transmission through scattering samples exhibits temporal distortion, as shown in the pixel-wise time evolution of the pulses and random modulation in the spatial field distributions (video 1). Field amplitude is plotted 4 times the original spatial field at $14.40$ ps, $15.00$ ps, and $18.90$ ps. **c.** The pixel-wise spectrum of scattered fields corresponds to the pulses shown in a. **d.** Mean free path of the scattering sample within the THz source bandwidth **e.** The spatial multiplexing of broadband THz fields and phases illustrates the broadband speckle synthesis within the THz source bandwidth (video 2).

**Scattering media characterisation using knife-edge wavefront shaping**
The core idea is to provide a unique scattering association to different shielding portions of the broadband terahertz field illumination (as per ref[45,46]) with a knife-edge moving in the near-field of the scatterer. The scatterer waveform obtained by the 2D time-domain detection set is used as a base to reconstruct a 1D transmission matrix. Once the transmission matrix is known, the image of any 1D object can be deterministically retrieved. Figure 2a illustrates a typical setting for a knife edge (a thin conductive blade clipping a Gaussian beam) scan, where the THz near-field imager collects scattered fields from the sample partially illuminated. We can write the spatiotemporal field profile ($E_s$) generated by the specific placement of the knife at the source plane as,

$$E_s(x, y, z_s; t) \propto \mathrm{K}(x, y_o, z_s) E_{in}(x, y, z_s; t) \qquad (3)$$

where $\mathrm{K}(x, y_o, z_s)$ represents the point-wise spatial response of the knife edge, which is defined as a one-dimensional Heaviside step function in the transverse direction with a fixed coordinate position $(y_o, z_s)$ and $E_{in}$ is the time-resolved Gaussian beam profile from the source. Initially, we characterised the spatial beam profile for the reference THz field (included in the supplementary figure S4) to normalise the results. The key idea here is that (figure 2a) because of the near-field coupling, the knife edge can selectively couple with a relatively large number of modes in the scattering volume (i.e. discernible scattering outputs) not necessarily accessible through far-field illuminations.

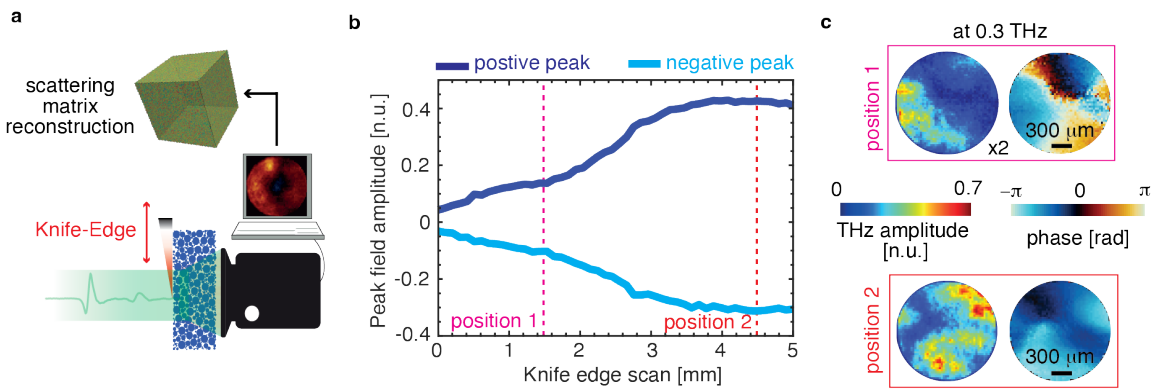

**Figure 2: Scattering media characterisation using knife edge wavefront shaping. a.** Schematics of the experimental setup showing a knife-edge shapes the impinging THz wavefront at the front end of the scattering medium (8% fractional mass concentration of Silicon microparticles embedded in paraffin) and corresponding scattered fields are collected using a THz time-domain imager placed within the near-field region. By combining the transmitted fields corresponding to the various knife edge positions, the space of potential 1D spatiotemporal images could be derived and further constructed as a hyperspectral combinatory scattering matrix to characterise the scattering sample. **b.** variation of pulse peak value with respect to the various transverse position of knife edge. **c.** spatial field amplitude and phase profile at 0.3 THz corresponding to knife edge position 1 and 2 (dashed lines shown in b).

Figure 2b illustrates the variation in positive and negative peak values of the pulse as a function of the transverse position of the knife edge. In the spectral domain, the spatial field amplitude and phase distribution at 0.3 THz, corresponding to the input wavefronts generated by two different knife-edge positions (at 1.5 mm and 4.5 mm), are shown in Figure 2c, highlighting the influence of the scattered field from the knife edge positions.

We can write fields at the source and detection plane as vectors $e_s \in \mathbb{C}^{S \times 1}$ and $e_d \in \mathbb{C}^{D \times 1}$, and define a scattering combinatory transfer matrix as $T_{ds} \in \mathbb{C}^{S \times D}$, for each frequency $\omega$. S and D represent the number of independent pixels (or segments) that sample the input and output field plane, respectively. For R number of independent spectral sampling points, $T$ is a $\mathbb{C}^{S \times D \times R}$ three-dimensional matrix, defined as $T_{ds}(\omega_r)$. For the given r-th spectral mode, the linear relationship between generic s-th source and d-th detection spatial segments reads as follows,

$$e_d(\omega_r) = \sum_S T_{ds}(\omega_r) e_s(\omega_r) \tag{4}$$

Further, for each i-th position of the knife edge, we collect the corresponding $e_d^{(i)}(\omega_r)$ and column-wise stack in a measurement matrix $M(\omega_r) \in \mathbb{C}^{D \times P}$ and corresponding synthesised THz wavefront in decomposition matrix $X(\omega_r) \in \mathbb{C}^{S \times P}$, where P is the number of wavefronts modulated by the edge of the knife. To identify the transfer matrix at frequency $\omega$, we set up a least square optimisation problem as,

$$T_r^\dagger(\omega_r) = \mathrm{argmin}_{T_r \in \mathbb{C}^{S \times D}} \frac{1}{2} \left\| M(\omega_r)^\dagger - X(\omega_r)^\dagger T_r^\dagger(\omega_r) \right\|_2^2 \tag{5}$$

where † represents matrix conjugate transpose operation. We solve such an optimisation problem using an active-set algorithm[47].

**Hyperspectral image retrieval of a hidden object behind the scattering media**
The ability to reconstruct combinatory transfer matrix of the scatterers allows for the imaging of objects obscured by a scattering medium. Figure 3a illustrates the implementation of our image reconstruction process, where a metallic object $O(x, y)$ is positioned on the input surface of scattering medium. This shapes the impinging THz field and results in convoluted spatiotemporal scattered field distribution at the output of the media as,

$$D(x', y'; t') = \iiint T_x(x, y; x', y'; t, t') O(x, y) E_{in}(x, y; t) dx dy dt \tag{6}$$

where $D(x', y', t)$ is spatiotemporal field distribution recorded by THz near field imager. To extract the information of a hidden object behind the complex media from the measurements, we conduct a standard deconvolution of the retrieved combinatory transfer matrix, which produces the time-resolved field image $E_{retreived}(x, y, t)$ as follows:

$$E_{retreived}(x, y; t) = \mathcal{F}^{-1}\left[\widetilde{Tr}(x, y; x', y'; \omega)^{-1} * \widetilde{D}(x', y'; \omega)\right] \tag{7}$$

where $\mathcal{F}^{-1}$ and * represents the inverse time-Fourier transform and spatial convolution operation, respectively, and $\widetilde{D}(x', y'; \omega)$ is the time-Fourier transform of the recorded signal. To perform deconvolution, the essential task involves determining the inverse of $\widetilde{Tr}(x, y; x', y'; \omega)$. We employed the Moore–Penrose pseudo-inversion method, implemented by a truncated singular-value decomposition. Figure 3a depicts the experimental arrangement that involves acquiring time-domain scattered fields by placing a one-dimensional image object (200 $\mu m$ copper wire) at the anterior end of the scattering sample. Initially, we perform a Time-Fourier transform of the recorded signal to extract the spectroscopic information available for each pixel. As illustrated in Figure 3a, it is evident that the transmitted field-phase profiles of the object, subsequent to exposure to a uniform THz pulse, undergo a complete scrambling process of information within the scattering sample. The scrambled information of the object is further employed with the direct deconvolution operation as delineated by equation (7) in order to conduct the image retrieval procedure. Intriguingly, the outcome of this reconstruction (figure 3b) is an image retrieval process across the extensive bandwidth of the THz spectrum. We can observe the remarkable matching with the measured direct field image without a scattering medium, in Figure 3c. Intriguingly, because the wire is metallic, we can appreciate that the post-scattered field polarised along the wire direction, x. In the spectrum domain, the phase image of the wire is clearly accompanied by lateral echoes that represent the phase delay accumulated by the scattered field along x.[48]

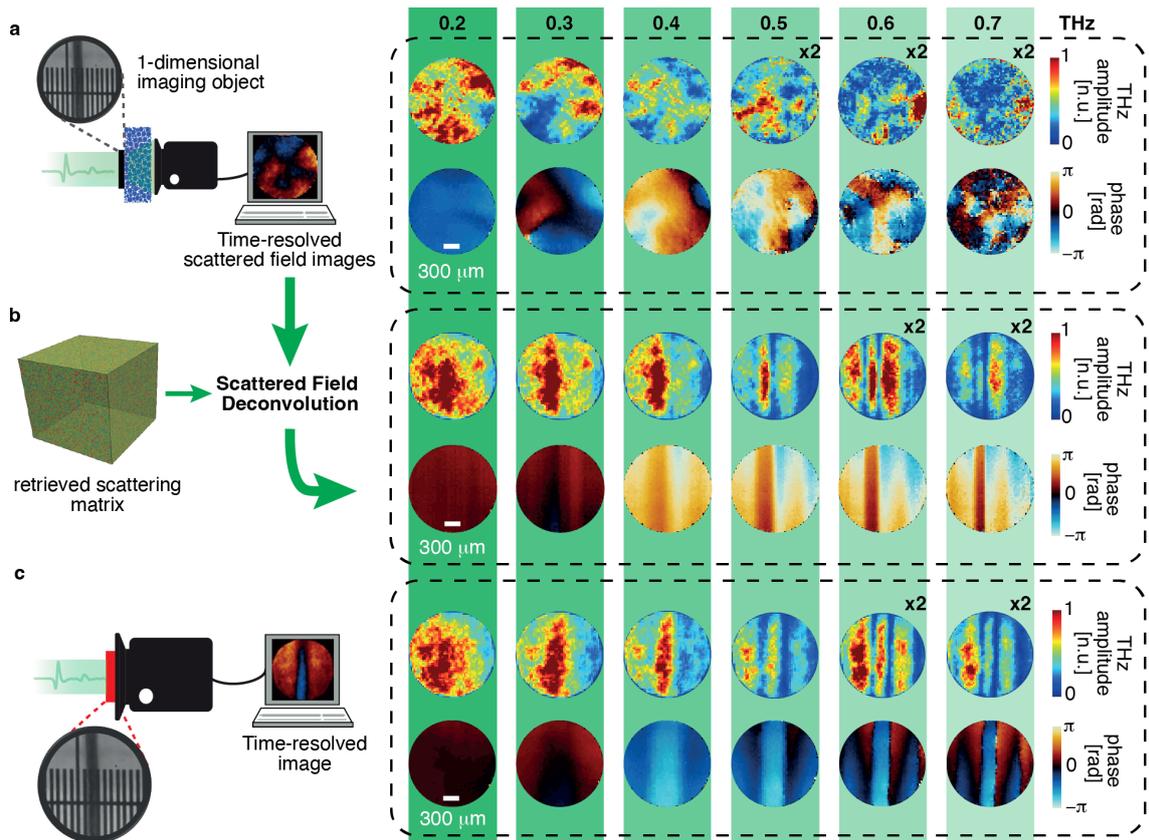

**Figure 3. Hyperspectral image retrieval through scattering media. a.** Broadband scrambled field and phase profiles of a 1D imaging object (200 $\mu m$ thick copper wire) hidden behind the scattering media (video 3). **b.** Hyperspectral image retrieval: the data in a. are deconvoluted by the reconstructed transmission matrix of a scattering sample (video 4). **c.** Hyperspectral THz images of the 1D imaging object used in a. for comparison (video 5).

## Methods
### Experimental settings and data post-processing

The experimental setting is based on a full-field Time-Domain Spectroscopy microscope as per ref [49,50]. A terahertz field generated by a large aperture $LiNbO_3$ prism, excited with a tilted front pulse from a regenerative amplified source, is projected into the detection crystal (a thin $LiNbO_3$) that functions as an imaging platform, as per supplementary S2. The field of view offered by THz imager is about 1.44 mm × 1.44 mm. The reflection of a large probe, impinging on the opposite side of the detection crystal performs a large-area electro-optic sampling, its relative delay being controlled by a delay stage placed on the optical THz pump. An imaging system project to orthogonal circular polarisation components of the probe beam onto a CMOS camera sensor, which results in a spatial resolution of 28.8 μm spatial period (obtained by binning the 600x600 image resolution output down to 50x50 to achieve appropriate signal-to-noise ratio). Supplementary information S3 delves into describing the role of detection plane spatial sampling. The typical performance of the imaging setup is shown in Figure S5, which serves as a benchmark for our imaging system. As we compare our results with the ongoing investigation in the optical domain, it is important to stress that the temporal coherence of scatterers is a pivotal problem at optical wavelength (i.e. the medium undergoes rapid thermal changes at scales comparable to or larger than the wavelength), this is significantly mitigated by the long terahertz wavelength. Hence, we did not experience any drift on the scattered pattern within the typical timeframe of an experimental session.

## Conclusion
We experimentally demonstrated a deterministic approach towards imaging through complex scattering media operating with broadband THz pulses. A large area-field detection is exploited to perform an orthogonal sampling decomposition of the scattering transmission matrix. Regardless of the relatively low density of scattering element in the medium, a near-field coupled knife-edge allows the excitation of a sufficiently large number of separable scattering modes to allow complete reconstruction of a field distribution through the medium. We performed a benchmark of this concept using a 1D image. As this approach does not exploit any statistical description of the scatterer and simply applies a super-position principle, we demonstrated full operation with broadband waveforms, which translates into the ability to reconstruct the full scattered spectrum from the object, i.e. their spectral fingerprint. This specific aspect is currently the subject of further studies. We believe our microscopy approach is a gateway to new possibilities in studying complex media, specifically scattering microstructures like cells or microorganisms.

**AUTHOR INFORMATION**
**Corresponding Author**
Marco Peccianti − Emergent Photonics Research Centre, Department of Physics, School of Science, Loughborough University, LE11 3TU, U.K.
email: m.peccianti@lboro.ac.uk.

**Author Contributions**
All authors were engaged in the general discussion regarding the conceptualisation. Experimental investigators: V.K. (lead), V.C. and A.C. All authors contributed to the general understanding of the results and drafting of the paper, the formal analysis and data curation. J.S.T.G., L.P., L.O. A.P. and M.P. supervised the research activities.

**Funding Sources**
This project received funding from the European Research Council (ERC) under the European Union's Horizon 2020 Research and Innovation Programme Grant No. 725046. This project has received funding from the European Research Council (ERC) under the European Union's Horizon 2020 research and innovation programme grant agreement no. 851758 (TELSCOMBE). The authors acknowledge financial support from the (UK) Engineering and Physical Sciences Research Council (EPSRC), Grant Nos. EP/X012689/1 and EP/S001018/1 and Leverhulme Trust (Early Career Fellowship ECF-2020-537 and Early Career Fellowship ECF-2022-710, Early Career Fellowship ECF-2023-315 Project grant RPG-2022-090), and from the Loughborough University Vice-Chancellor Independent Research Fellowship (STARLIGHT).

**Data Availability Statement**
The data that support the findings of this study will be openly available in https://repository.lboro.ac.uk/ upon publication.

# Supplementary materials

## Terahertz Microscopy Through Complex Media

Vivek Kumar[1,3], Vittorio Cecconi[1,2], Antonio Cutrona[1,2], Luke Peters[1,2], Luana Olivieri[1,2], Juan S. Totero Gongora[1,2], Alessia Pasquazi[1,2], Marco Peccianti[1,2]*

[1]Emergent Photonics Lab (EPic), Department of Physics and Astronomy, University of Sussex, BN1 9QH, UK
[2]Emergent Photonics Research Centre, Department of Physics, School of Science, Loughborough University, LE11 3TU, UK
[3]Laboratoire Kastler Brossel, ENS-Universite PSL, CNRS, Sorbonne Universite, College de France, 24 rue Lhomond, 75005 Paris, France

*m.peccianti@lboro.ac.uk

The Supplementary Information comprises 6 pages and 5 figures.

### Supplementary note S1
#### S1.1-Scattering sample information

Our typical scattering sample used in the experiment is fabricated by embedding Hi-Z Silicon microparticles (150 µm – 300 µm from Ferroglobe) in a paraffin medium (shown in Fig S1(A)). Interestingly, the refractive index of Si (n=3.4) is largely independent of the frequency throughout the spectral range of the measurements. The embedding material, paraffin wax, is fully transparent at THz frequencies and has a specific melting point between 46-68℃, density of ~0.9 g/cm$^3$ that offers low thermal conductivity and a high heat capacity. Hence it is practical as a host material. Hi-Z silicon does not offer significant carrier-driven absorption at the spectral range used; hence multiple scattering is solely responsible for field modulation.

Initially, we performed a series of TDS experiments to characterise scattering samples of varying thickness and scatterers mass density by measuring the transmission of coherent, single-cycle THz pulses. Further we choose scattering sample that consists of 8% fractional mass concentration for Si particles with total sample thickness of 5.53 mm.

#### S2.2- Terahertz characterisation of the scattering sample

The combined study of concepts and methodologies from the field of complex wave propagation in scattering media and THz-TDS was reported early in 2000[1,2]. Several other groups have been investigated terahertz radiation transport phenomena in ensembles of dielectric spheres[3], subwavelength size metallic particles[4], granular composite materials[5,6] and random assemblies of spherical silica (SiO$_2$) particles in a paraffin matrix[7]. We performed similar TDS experiments to characterise scattering samples by measuring the transmission of coherent, single-cycle THz pulses as shown in Figure S1 (B-C). The transmitted wavefronts allow extracting the mean free path, $l_s(\omega)$, of the sample over a broad bandwidth by measuring the transfer function of a sample[2,7]. To this end, we first collected both reference and sample waveform and obtained transfer function for a sample as,

$$H_{sample}(\omega) = \frac{E_{sample}(\omega)}{E_{ref}(\omega)} \tag{S1}$$

where, $E_{sample}(\omega)$ and $E_{ref}(\omega)$ are time-Fourier transform of terahertz field profile propagated through a sample and free space, respectively. Further, scattering mean free path of a scattering sample over the entire bandwidth can be obtained as,

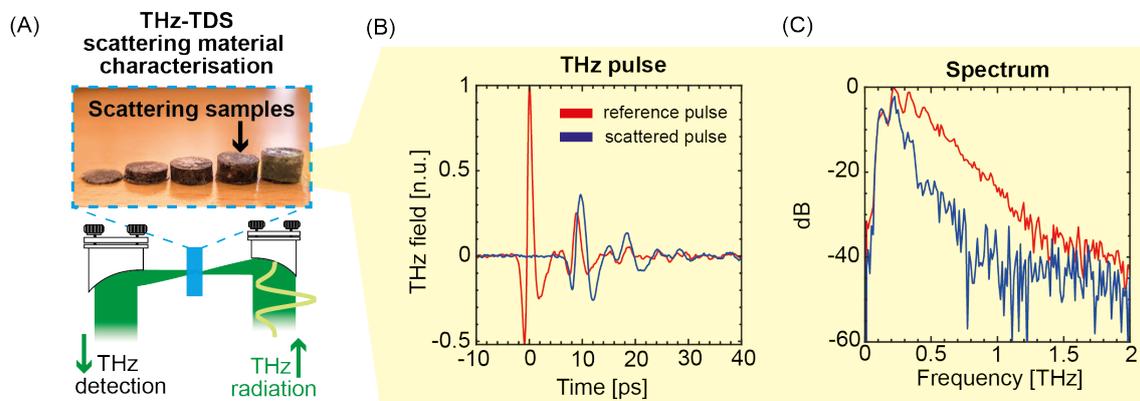

**Figure S1: Scattering sample characterisation. (A)** Schematic of THz TDS in Transmission Configuration **(B)** THz-TDSpulses for reference and sample. **(C)** Associated spectrum for the reference and scattered pulse.

$$l_s(\omega) = -\frac{L_0}{2\ ln(H(\omega))} \qquad (S2)$$

where, $L_0$ is the thickness of scattering sample.

**Supplementary note S2- Experimental setup**
The experimental setting employs a regenerative amplifier Ti: Sapphire laser system to provide an 1mJ optical pulse with a repetition rate of 1 kHz and a pulse duration of 90 fs at a central wavelength of 800 nm. As shown in Fig. S2, a beam sampler divides the laser into two beams, i.e., pump and probe. The pump beam is delayed in time with a motorised stage and undergoes front-tilt correction with the combined used of a diffraction grating (1800 lines/mm) and lenses (L1=250 mm, L2=150 mm). The pump pulse is nonlinearly converted into a THz beam via optical rectification on the Cherenkov angle of a stochiometric MgO (0.6%) prism cut Lithium Niobate crystal[8–11]. The crystal hosts an optical rectification process that exhibits maximum phase-matching also along the normal of the output facet[11]. The very large interaction length, the relatively high nonlinear susceptibility and the damage threshold of the Lithium Niobate prism allow for remarkably high optical to THz conversion efficiency (0.1% and above). Further, the THz pulse is modulated via a 10 Hz chopper obtained as an exact division of the laser repetition rate, and then it illuminates an arbitrary imaging sample. On the other hand, the probe beam is imaged on the back side of a quadratic detection crystal (a 20 $\mu m$ - Lithium Niobate film deposited on a glass substrate) utilising a telescopic combination of lenses L3 (300 mm), L4 (100 mm) and L5 (100 mm), L6 (100 mm). The facet at the other end of the detection crystal is coated with a highly reflective coating. The result is that the reflected probe co-propagates with the impinging THz wave, realising a large spatial area of electro-optical sampling, thanks to the THz-induced Pockels effect. Hence, the spatial distribution of the polarisation state of the reflected probe beam carriers the image of the detected THz field. Following a standard approach in practice, a large quarter wave plate, combination of lenes (L7=75 mm, L8=150 mm) and a polarising beam splitter separate the two images of the clockwise and anti-clockwise circular polarised components that are both projected onto the same CMOS camera (DMK23UP1300)[12–14].

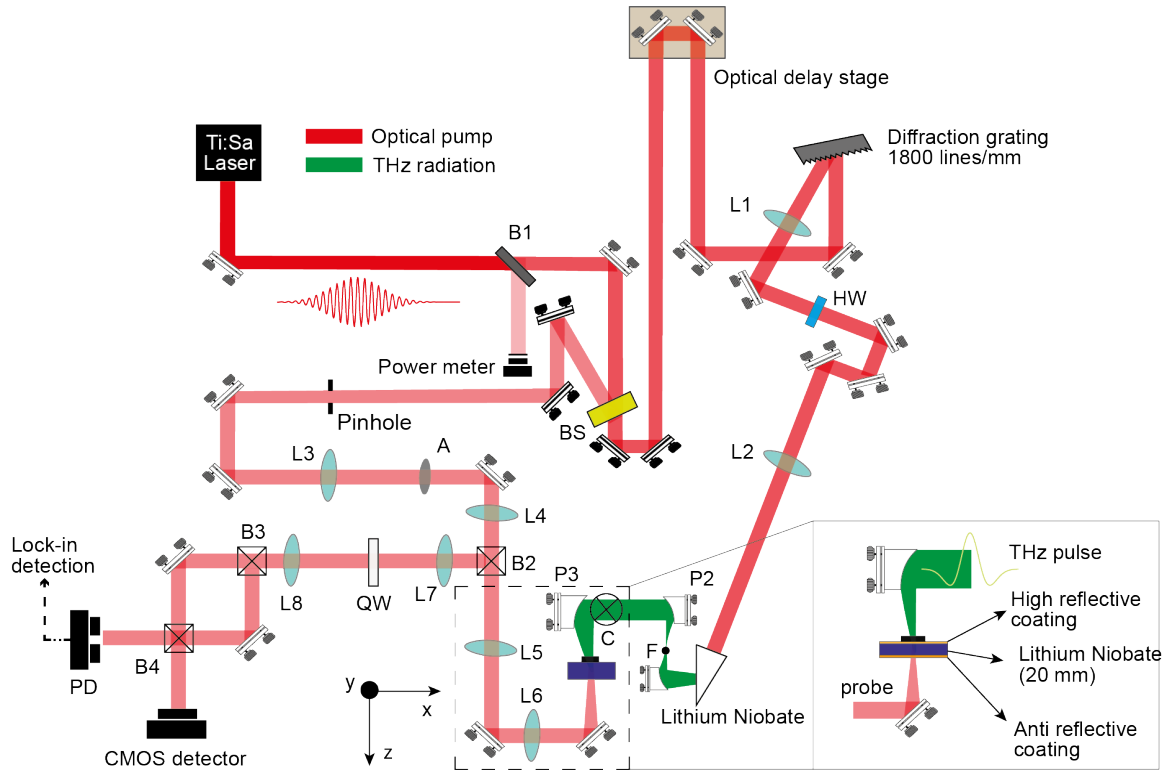

**Figure S2: Schematic of an experimental setup for terahertz near field imager.** B1 – 99:1 beam splitter, B2, B4 – 50:50 beam splitter, B3 – 50:50 polarised beam splitter, BS – Beam sampler, C – chopper, HW – half waveplate, PD – balanced photodiodes, A – attenuator, F – filter, QW – quarter waveplate, L(1-8) – plano-convex lens (L1=250 mm, L2=150 mm, L3=300 mm, L4=100 mm, L5=100 mm, L6=100 mm, L7=75 mm, L8=150 mm), P – Parabolic mirrors (P1=25.4 mm, P2=76.2 mm, P3=50.3 mm).

**Supplementary note S3- The role of detection plane spatial sampling**

A fundamental aspect in assessing the scattered field is related to the broadening of the output spatial spectrum induced by scattering. This is particularly relevant when we consider a near-field detection, where transverse spatial-frequency components can have sub-wavelength periods, i.e. momentum exceeding the vacuum propagation constant. From this argument, it descends that the output spatial sampling period during the image acquisition (in the two transverse dimensions) should be significantly smaller than the wavelength. Besides, the CMOS array sampling period (4.8 μm) results in the resolving power of the imaging system (better than 7 μm ) orders of magnitude smaller than the peak terahertz wavelength and largely within the optical diffraction limit of the probe assembly. Conversely, large detection areas correspond to higher detected signals and lower signal-to-noise ratio. Figure S3(A) and S3(B) present the typical benefit obtained by binning together the camera element to obtain a macro-pixel of a given size. In terms of procedure, then, the output sampling will be chosen, analysing the output spatial spectrum at different scales, determining $\Delta C$ from the roll-off of the spatial spectrum at high frequency, and keeping the spatial sampling $\Delta_x$ one order of magnitude smaller than $\Delta C$ (corresponds to a Nyquist frequency Limit of $5/\Delta C$).

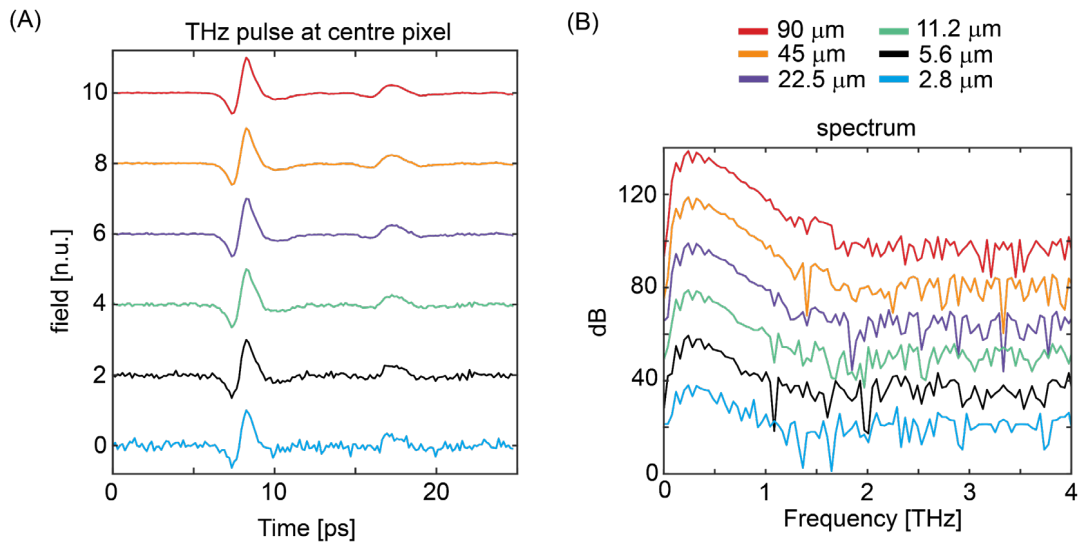

**Figure S3: Spatial sampling of detection plane. (A-B)** Signal-to-noise variation in THz pulse and its spectrum collected at a centre macro pixel by varying pixel size.

**Supplementary figures**

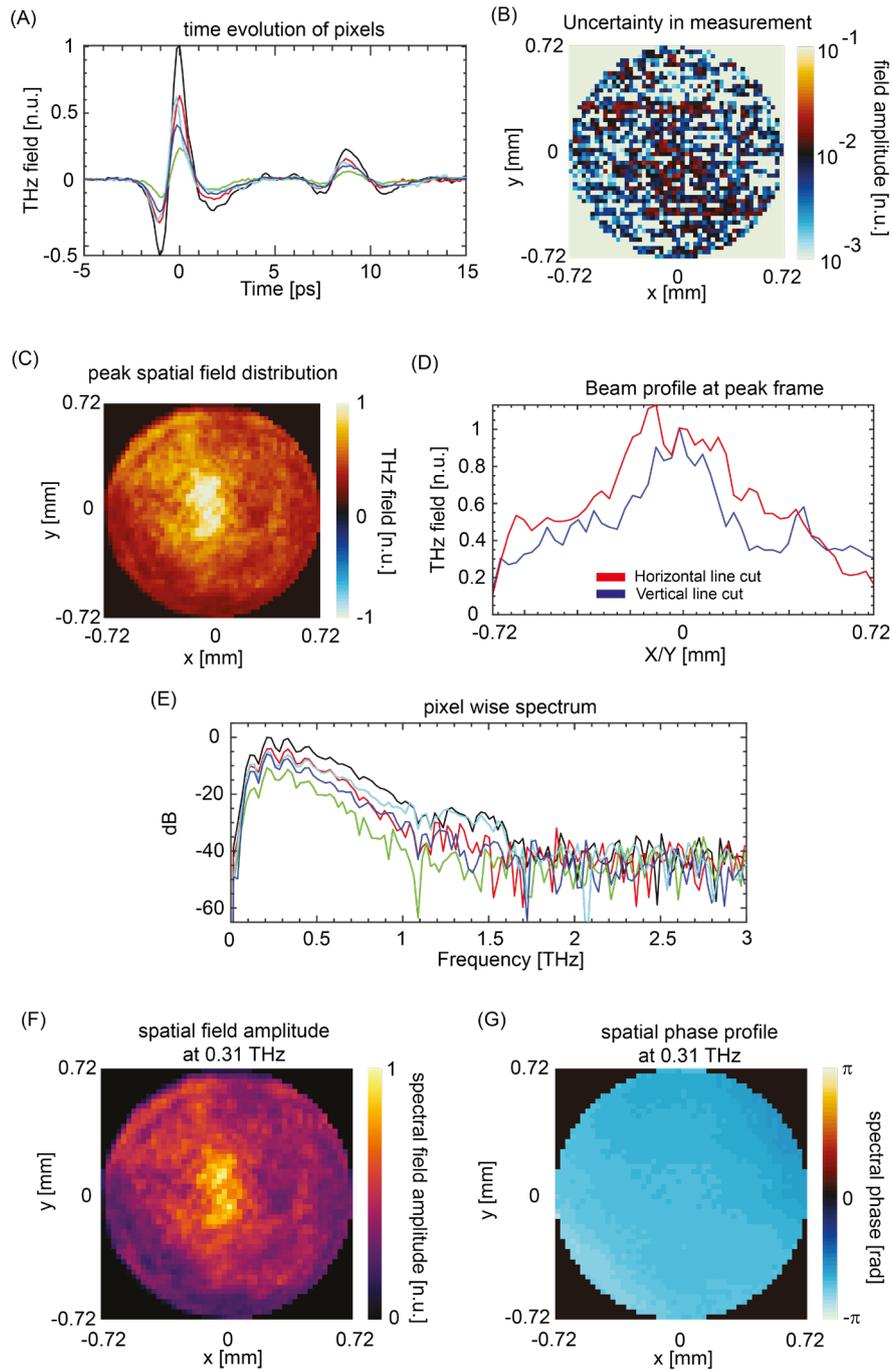

**Figure S4: Time-resolved reference beam profile obtained from terahertz near-field imager.** **(A)** Time-evolution of some exemplary pixels. **(B)** Uncertainty (noise) map in spatial field measurements. **(C)** Spatial field distribution at 0 ps. **(D)** Beam profile obtained for spatial field distribution at the peak of pulse. **(E)** Terahertz reference spectrum corresponding to various pixels. **(F)** Spatial field amplitude profile at 0.31 THz. **(G)** Spatial phase distribution at 0.31 THz.

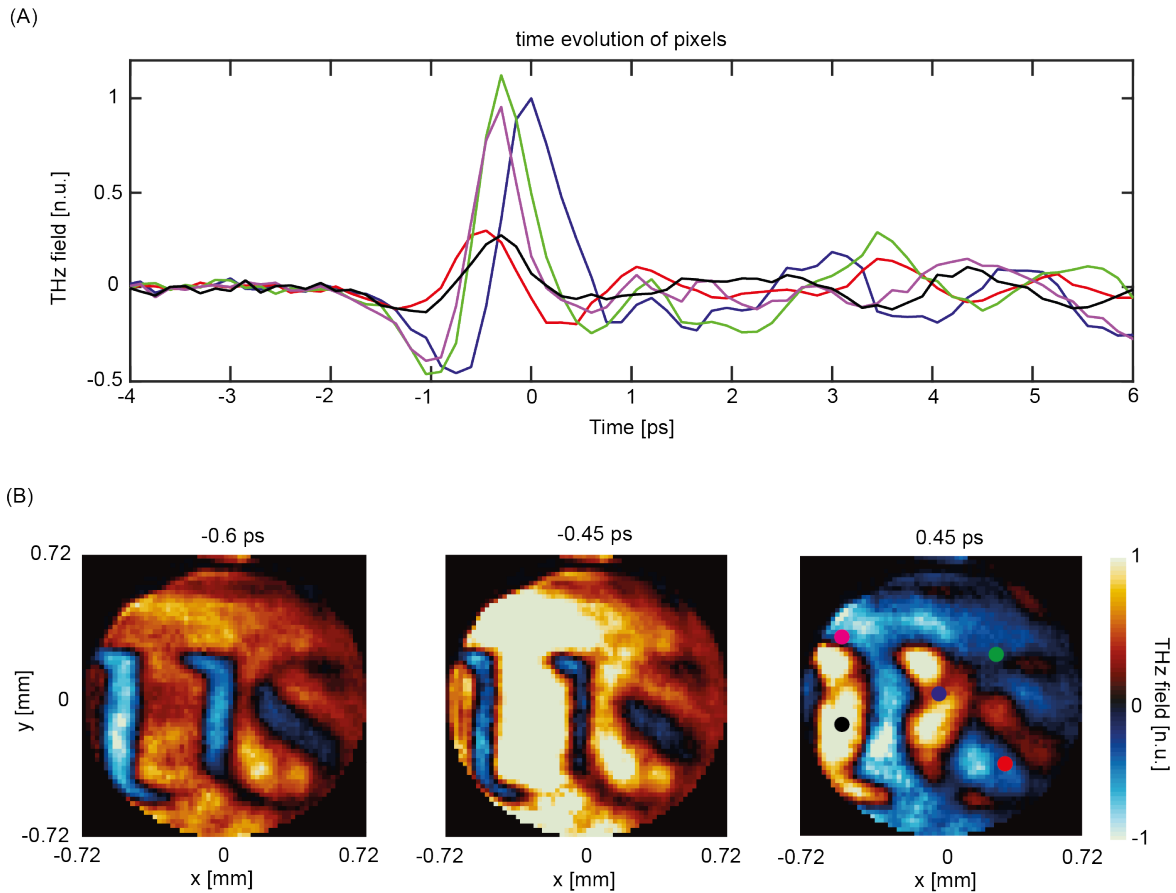

**Figure S5: Time-resolved image of metallic mask collected by terahertz near field imager.** (A) Time evolution of various pixels. (B) Spatial field distribution of metallic mask at -0.6 ps, -0.45 ps and 0.45 ps.